\documentclass[twocolumn,twocolappendix  ]{aastex631}
\usepackage{amsmath}

\shorttitle{Microlensing of Technosignatures}
\shortauthors{Baghram}

\begin{document}
	
\title{Microlensing Signatures of Dyson Sphere\textendash like Structures around Primordial Black Holes as Technosignatures of Extraterrestrial Advanced Civilizations}

	\author[0000-0001-6131-4167]{Shant \surname{Baghram}}
	\email{baghram@sharif.edu}
	\affiliation{Department of Physics, Sharif University of Technology, Tehran 11155-9161, Iran}
	\affiliation{Research Center for High Energy Physics, Department of Physics, Sharif University of Technology, Tehran 11155-9161, Iran}

	
\begin{abstract}
We investigate the microlensing detectability of extraterrestrial technosignatures originating from Dyson sphere–like structures, such as Dyson Swarms surrounding primordial black holes (PBHs). These hypothetical swarms consist of stochastically varying, partially opaque structures that could modulate standard microlensing light curves through time-dependent transmission effects. We introduce a probabilistic framework that includes a stochastic transmission model governed by variable optical depth and random gap distributions. We perform a parameter scan and generate heatmaps of the optical transit duration. We study the infrared excess radiation and peak emission wavelength as complementary observational signatures. Additionally, we define and analyze the effective optical depth and the anomalous microlensing event rate for these stochastic structures. Our findings provide a new avenue for searching for extraterrestrial advanced civilizations by extending microlensing studies to include artificial, dynamic modulation signatures.
\end{abstract}

	\keywords{ \object{Astrobiology (74)},
		\object{Search for extraterrestrial intelligence (2127)}, \object{ Primordial black holes (1292)}}
\section{Introduction}
\label{Sec1}
The search for extraterrestrial intelligence (SETI), a continuation of the multidisciplinary quest known as astrobiology \citep{Plaxco:2001}, has traditionally focused on detecting intentional signals, such as radio or laser transmissions  \citep{1959Natur.184..844C,Drake1961,Tarter:2001kq,2016ApJ...828...19R,anderson2025seti}. This raises the question of other complementary fingerprints of their presence. Extraterrestrial intelligence may also evolve into an advanced civilization that constructs and develops megastructures around its host star to harvest energy \citep{Tegmark:2017}. The Dyson spheres are one of the known ideas for advanced civilizations in this direction \citep{Dyson1960a,Dyson1960b} (for a recent review see \cite{2020SerAJ.200....1W}). This extraordinary idea has been invoked a few times to explain inexplicable signals observed in various studies.~A notable example is the study of the Boyajian star \citep{2016MNRAS.457.3988B}. Accordingly, now a complementary strategy has emerged: the search for \emph{technosignatures}, indirect indicators of advanced civilizations inferred through their engineering impact on their environment \citep{Gajjar:2019rvn,Green:2020jor}. Dyson spheres\textemdash vast artificial structures designed to harness energy from stars or nearby compact objects\textemdash remain among the most compelling hypothetical technosignatures \citep{2025MNRAS.537.1249M} and ideas are proposed for their detection \citep{2025MNRAS.538L..56R}.
In a recent work by \cite{2025ApJ...978..132B}, a novel idea is proposed, in which an advanced civilization used primordial black holes (PBHs) \citep{1966AZh....43..758Z} and their accretion disk as the source of energy. Recent constraints on the abundance of compact dark matter (DM) in the form of PBHs have revived interest in detecting these objects via gravitational microlensing \citep{Carr:2020gox}. PBHs are prominent candidates for DM \citep{Green:2020jor}, and gravitational waves (GW) detected by GW interferometer detectors \citep{Bird:2016dcv}. \\In this work, we propose combining these lines of inquiry: using gravitational microlensing not only to detect PBHs, but also to identify anomalous lensing events consistent with the presence of surrounding Dyson sphere\textendash like megastructures.
If such a structure exists, it would modify the microlensing signature of the PBH in characteristic ways. These include partial suppression of optical light due to obstruction, infrared excess from thermal reradiation, and possibly chromatic deviations from the standard achromatic lensing profile.
We aim to characterize these anomalies and propose strategies for detecting them in current and future microlensing surveys. In this direction, we study the microlensing effect of the Dyson Swarm as a plausible candidate for advanced civilizations' technology. A Dyson Swarm is a conceptual megastructure made up of countless satellites or habitats that orbit a star, capturing and using its energy. Unlike the better-known Dyson sphere—a solid shell that completely encircles a star—a Dyson Swarm provides greater flexibility in how it is built and maintained \citep{smkith2022}. This option is generally regarded as a more practical and achievable way to harness the energy of a star. 
The structure of this work is as follows: In Section \ref{Sec2}, we review the concept of Dyson sphere\textendash like structures around PBH. In Section \ref{Sec3}, we review the fundamentals of microlensing from a point source and the surrounding Dyson sphere\textendash like structure around PBH. In Section \ref{Sec4}, we demonstrate our results and the observational strategy, and finally in Section \ref{Sec5}, we conclude and have our future remarks.
\section{Dyson sphere\textendash like structure around PBH}
\label{Sec2}	
Advanced civilizations can be categorized by the usage of energy: The Kardashev scale is a method of measuring a civilization’s level of technological advancement based on the amount of energy it can harness and use \citep{Kardashev1964}. Proposed by Soviet astrophysicist Nikolai Semyonovich Kardashev in 1964\footnote{At a conference in Armenia, Byurakan observatory, he presented a paper titled "Transmission of Information by Extraterrestrial Civilizations".}, the scale envisions three classic types: Type I harnesses all the energy available on its home planet, Type II uses the total energy of its star (often imagined with a megastructure like a Dyson sphere), and Type III utilizes the energy of its entire galaxy. The concept provides a framework for thinking about the long-term goals of civilizations and the energy requirements. Since Kardashev’s original formulation, many refinements and continuous scales have been proposed to account for partial energy. Still, the core idea remains a useful heuristic for gauging extraterrestrial civilizations and the possible trajectories of our own technological development.
Advanced civilizations, which are classified as Kardashev scale types two and higher, require an enormous amount of energy \citep{Kardashev1964}. 
The concept of a Dyson sphere, proposed by Freeman Dyson, describes a structure designed to harvest the entire energy output of a host star of an exoplanet dominated by an advanced civilization \citep{Dyson1960a,Dyson1960b}.
With a simple thermal equilibrium assumption, the temperature of the Dyson sphere appears to depend only on its radius. {{However, in a more realistic scenario, part of the energy would be used by the civilization, and the Dyson Sphere would re-emit the waste heat at a different effective (blackbody) temperature. So there are two temperatures, one is the "inner/environment" temperature and the other is the "outer/waste heat" temperature $T_{\rm{waste}}$. In order to formulate this statement, we set the relation of the source luminosity $L_{\rm{source}}$, which provides the advanced civilizations energy, to the waste heat temperature as
		\begin{equation}
			(1-\epsilon)f_{\rm{cap}}L_{\rm{source}} = 4\pi R^2_{\rm{Dyson}}\Upsilon\sigma_{\rm{SB}}T_{\rm{waste}}^4,
		\end{equation}
where $R_{\rm{Dyson}}$ is the radius of Dyson sphere\textendash like structure. $f_{\rm{cap}}$, and $\Upsilon$ are capture fraction and radiating fraction of the structure.~$\sigma_{\rm{SB}}\simeq5.67\times 10^{-8}~ \rm{W m^{-2} K^{-4}}$ is Stefan–Boltzmann constant. ~$\epsilon$ is energy transfer efficiency factor and $1-\epsilon$ is the waste energy fraction. For simplicity of the argument, in this work, we set $\epsilon\simeq 0.2$ and we set $f_{\rm{cap}}=\Upsilon$. Hereafter, the waste heat temperature assigned to the Dyson sphere\textendash like structure's temperature ($T_{\rm{Dyson}}=T_{\rm{waste}}$). \\ 
Accordingly, the equilibrium temperature of the Dyson sphere depends on its radius.
The radius of this megastructure around a star, denoted as $R_{\rm{Dyson}}|_{\rm{star}}$, can be approximated as 
\begin{equation}
	R_{\rm{Dyson}}|_{\rm{star}} \sim (1-\epsilon)^{\frac{1}{2}}  R_{\odot} \left(\frac{T_{\odot}}{T_{\rm{Dyson}}}\right)^2 ,
\end{equation}
where $T_{\rm{Dyson}}$ is the temperature assigned to waste heat of the Dyson sphere\textendash like structure as mentioned above, $R_{\odot}\sim 7\times 10^5 ~{\rm{km}}$ and $T_{\odot}\sim 5770 ~\rm{K}$ are radius and temperature of Sun, respectively.}}
This megastructure heats and reradiates in infrared \citep{2014ApJ...792...26W}. \\
{{Besides the host star, the advanced civilization can use the PBH within the limit of space exploration distance for energy harvesting \citep{2025ApJ...978..132B}}. 
\cite{Hsiao:2021qij} showed that the main contribution of a black hole as an energy supplier comes from its accretion disk.
For a black hole with mass $M$, the Eddington luminosity $L_{\rm{Edd}}$ is as below \citep{Joss1973,Rybicki1979}.
\begin{equation}
	L_{\rm{Edd}} = \frac{4\pi GMm_p c}{\sigma_T}\sim 3.2 \times 10^4 (\frac{M}{M_{\odot}})L_{\odot},
\end{equation}
where $m_p$ is the mass of proton, $G$ is the Newton constant, $c$ is the speed of light, $\sigma_T=6.65\times 10^{-29} m^{-2}$ is Thomson scattering cross section. $M_{\odot}\sim 2.0 \times 10^{30} ~\rm{kg}$ and $L_{\odot}\sim 3.8\times 10^{26}~ \rm{W}$ are the mass and the luminosity of Sun. The luminosity of disk $L_{\rm{disk}}$ with mass loss rate of $\frac{dm}{dt}$ is
\begin{equation}
	L_{\rm{disk}}=\eta_{\rm{disk}}\frac{dm}{dt}c^2, 
\end{equation}
where $\eta_{\rm{disk}}$ is the efficiency of  accretion disk\footnote{The assumption that an accretion disk can form around PBHs is discussed in several studies  such as \cite{2008ApJ...680..829R, 2017PhRvD..95d3534A,2018MNRAS.479..315S,2020A&A...642L...6B}. The efficiency of accretion of different type of black holes is different. The mass accretion rate $dm/dt$ varies across different AGNs/accretion disks (even for the same black hole mass).}. \\
{{Now the accretion disk of PBH can act as a source $	L_{\rm{source}} = L_{\rm{disk}}$.
In this scenario, around a PBH with mass $M_{\rm{PBH}}$, one can make Dyson sphere\textendash like structures with a waste energy fraction of $(1-\epsilon)$ and a corresponding temperature $T_{\rm{Dyson}}$ and radius $R_{\rm{Dyson}}|_{\rm{\tiny{PBH}}}$ as
\begin{equation}
	(1-\epsilon)f_{\rm{cap}}\eta L_{\rm{Edd}} = 4\pi 	(R^2_{\rm{Dyson}}|_{\rm{\tiny{PBH}}})\Upsilon  \sigma_{\rm{SB}} {T^4_{\rm{Dyson}}},
\end{equation}
where $\eta\equiv \frac{L_{\rm{disk}}}{L_{\rm{Edd}}}$ is the ratio of bolometric luminosity to its Eddington luminosity (i.e. Eddington ratio). With the assumption of equality of capturing and radiating surfaces $f_{\rm{cap}}=\Upsilon$,  one can make Dyson sphere\textendash like structures closer due to small size of accretion disk,  approximately at a distance of $R_{\rm{Dyson}}|_{\rm{\tiny{PBH}}}$ as \citep{2025ApJ...978..132B}}}
\begin{equation} 
	R_{\rm{Dyson}}|_{\rm{\tiny{PBH}}}\sim 2.7 \times 10^7 {\eta^{\frac{1}{2}}} (1-\epsilon)^{\frac{1}{2}} (\frac{M_{\rm{PBH}}}{M_{\odot}})^{\frac{1}{2}}(\frac{\rm{K}}{T_{\rm{Dyson}}})^2  \rm{au}.
	\label{eq:Rdyson-pbh}	
\end{equation}
In this work, with a conservative assumption, we assume that the ratio of  bolometric luminosity to its Eddington luminosity (i.e. Eddington ratio) is  $\eta = 10^{-4}$, which is an order of magnitude more conservative assumption than the ratio suggested by \cite{Hsiao:2021qij}.
For the Eddington ratio of $\eta=10^{-4}$, efficiency of $\epsilon=0.2$, and a solar mass PBH, we will have a series of Dyson sphere radii as presented in Table \ref{Table1}. We list the distance of the  Dyson sphere\textendash like structure from a star or a PBH corresponding to the Dyson sphere\textendash like structures' temperature.
\begin{table}
	\begin{center}
		\begin{tabular}{ c | c | c |c }
			Dyson Temp. $T_{\rm{DS}}$ & $R_{\rm{Dyson}}|_{\rm{star}}$& $R_{\rm{Dyson}}|_{\rm{\tiny{PBH}}}$ & $\lambda_{\rm{max}}$\\ 
			\hline \hline
			3000 K & $\sim0.015$ au  &  $\sim0.03$ au & $\sim 966 ~ {nm}$  \\ 
			300 K & $\sim1.52$ au  & $\sim 2.68$ au  & $\sim 9.66 ~ \mu m$ \\
			30 K & $\sim 152$ au &  $\sim 268$ au  & $\sim 96.6 ~\mu m$\\ 
		\end{tabular}
		\caption{A table showing the Dyson sphere\textendash like structures' temperature (first column) and their distances from the energy source (star and PBH) in the second and third columns. The fourth column shows the peak of the wavelength in black-body radiation from a Dyson sphere\textendash like structure. $R_{\rm{Dyson}}|_{\rm{\tiny{PBH}}}$ is calculated for solar mass PBH with $\eta = 10^{-4}$. The efficiency of Dyson sphere\textendash like structure in both cases is set to $\epsilon=0.2$. Note that in table and figures $T_{\rm{DS}}$ is used for $T_{\rm{Dyson}}$ and $R_{\rm{DS}}$ for $R_{\rm{Dyson}}|_{\rm{\tiny{PBH}}}$,  interchangeably.}
		\label{Table1}
	\end{center}
\end{table}
\section{Microlensing by Primordial Black Holes and Dyson Sphere\textendash like structure}
\label{Sec3}
In this section, we discuss the microlensing of a compact single lens \citep{2008arXiv0811.0441M,rahvar2015gravitational,Dodelson2017}. Then assert that this single compact lens could be a PBH with a Dyson sphere\textendash like structure which can be used by an advanced civilization. Accordingly, we will also study the effect of the modified lensing signal.
\subsection{Standard lensing}
For a point mass lens, the lensing equation is
\begin{equation}
	\theta^2=\beta\theta+\theta^2_{\rm{E}},
\end{equation}
where $\theta$ is the angular position of the image as seen by the observer. $\beta$ is the original angle of the source and $\theta_{\rm{E}}$ is the Einstein angle for a lens with mass $M_l$ in a flat universe as 
\begin{equation}
	\theta_{\rm{E}} = \sqrt{\frac{4GM_l}{c^2}\frac{D_{ls}}{D_lD_s}} = \sqrt{\frac{2R_{\rm{sch}}}{D_s}\frac{1-x}{x}},
\end{equation}
where $D_l$, $D_s$ are distances to the lens and source, respectively. $D_{ls}$ is the distance of the lens to the source. $R_{\rm{sch}}$ is the Schwarzschild radius of the lens and $x\equiv D_l/D_s$.
The Einstein radius $R_{\rm{E}}=D_l\theta_{\rm{E}}$ will be
\begin{equation}
	R_{\rm{E}} = \sqrt{2R_{\rm{sch}}D_l(1-x)}.
\end{equation}
The magnification $A$ is given by the ratio of the image area to the source area
\begin{equation}
	A=|\frac{\theta d\theta}{\beta d\beta}| = \frac{\beta^2+2\theta_{\rm{E}}^2}{\beta(\beta^2+4\theta_{\rm{E}}^2)^{1/2}}.
\end{equation}
For a point-mass lens, the magnification of a source star is given by \cite{1986ApJ...304....1P}
\begin{equation}
	A(u) = \frac{u^2 + 2}{u \sqrt{u^2 + 4}},
\end{equation}
where $u \equiv\beta / \theta_{\rm{E}}$ is the normalized source angle to the Einstein angle. \\
Assuming a constant relative velocity between the lens and the source, and neglecting any acceleration, the angular position of the source—expressed in terms of the impact parameter—can be represented as the sum of the minimum impact parameter $\beta_0$ and the product of the relative angular velocity, $\mu$ and time.
\begin{equation}
	\beta^2=\beta_0^2+\mu^2(t-t_0)^2,
\end{equation}
where $t_0$ is the time of closest approach between the lens and the source and $\mu$ is defined as \citep{rahvar2015gravitational}
\begin{equation}
	\mu = \frac{{\bf{v}}_{s,\perp}-{\bf{v}}_{o,\perp}}{D_s} - \frac{{\bf{v}}_{l,\perp}-{\bf{v}}_{o,\perp}}{D_l},
\end{equation}
where ${\bf{v}}_{s,\perp}$, ${\bf{v}}_{l,\perp}$, ${\bf{v}}_{o,\perp}$ are perpendicular velocities of source, lens and observer to the line of sight. 
For the standard lensing signal, we rewrite the normalized $u$ as below:
\begin{equation}
	u^2(t)=u_0^2+\frac{(t-t_0)^2}{t^2_{\rm{E}}},
\end{equation}
where $t_{\rm{E}}=R_{\rm{E}}/v_{\rm{T}}=\theta_{\rm{E}}/\mu$ is the Einstein crossing time. $u_0$ is the minimum impact parameter, and $t_0$ is the time of maximum magnification. Note that $v_{\rm{T}}$ is the relative velocity of the lens and source.
An important quantity to introduce here is the optical depth \citep{rahvar2015gravitational}. It is defined as the probability that a given source is inside the Einstein ring of a lens at any given time \citep{2025ApJ...979..123N}
\begin{equation}\label{eq:tauDs}
	\tau(D_s)=\int_0^{D_s} \left[\int n(M_l,D_l)\sigma_{\rm{E}}(M_l,D_l)dM_l\right]dD_l,
\end{equation}
where $n(M_l,D_l)$ is the number density of the lenses in the mass range of $(M_l,M_l+dM_l)$ and the distance $D_l$ from the observer, and $\sigma_{\rm{E}}\equiv \pi R_{\rm{E}}^2$ is the cross section of lensing. Note that the optical depth defined in Equation (\ref{eq:tauDs}) is a function of the distance to the source. Now, we define the mass density profile of the lenses $\rho(D_l)$ as follows
\begin{equation}
	\rho(D_l)=\int  n(M_l,D_l)M_ldM_l.
\end{equation}
Then the optical depth will be
\begin{equation}\label{eq:tau-rho}
	\tau(D_s)=\pi\frac{4G}{c^2}\int_0^{D_s}\rho(D_l)D_l(1-x)dD_l.
\end{equation}
We will calculate this quantity for our configuration in Section \ref{Sec4}. The observable quantity is the integrated optical depth averaged over all detectable sources
\begin{equation}
	\tau=\frac{1}{N_s}\int_0^\infty\tau(D_s)N(D_s)dD_s,
\end{equation}
where $N(D_s)$ is the number density of sources at distance $D_s$ and $N_s\equiv \int_0^\infty N(D_s)dD_s$ is the number of all detectable source stars in the direction of sight.
\subsection{Modelling Dyson Sphere Induced Anomalies}
In this subsection, we assume that the source PBH of the advanced civilization has a Dyson sphere\textendash like structure. In the first step, to study the effect of this artificial megastructure on the lensing effect, we should compare the corresponding scales.
The Einstein angle must be compared with the angular size of the Dyson sphere\textendash like structure around a PBH to study the modified microlensing effect due to the Dyson sphere. So we define a specific angle $\theta_{\rm{Dyson}}|_{\rm{\tiny{PBH}}}$ radius as below:
\begin{equation}
	\theta_{\rm{Dyson}}|_{\rm{\tiny{PBH}}}=\frac{R_{\rm{Dyson}}|_{\rm{\tiny{PBH}}}}{D_l}.
\end{equation}
Now, one can define the corresponding ratio concerning the Einstein angle as
\begin{eqnarray}
	&&\Xi_{{\rm{D|PBH}}/{\rm{E}}} \equiv \frac{	\theta_{\rm{Dyson}}|_{\rm{\tiny{PBH}}}}{\theta_{\rm{E}}}\equiv \frac{R_{\rm{Dyson}|\rm{PBH}}}{R_{\rm{E}}} \\ \nonumber &&\simeq 9.5 \times 10^6 \eta^{\frac{1}{2}} (1-\epsilon)^{\frac{1}{2}} \left(\frac{\rm{K}}{T}\right)^2 \left(\frac{1}{1-x}\right)^{\frac{1}{2}}\left(\frac{\rm{kpc}}{D_l}\right)^{\frac{1}{2}}.
\end{eqnarray}
$\Xi_{\rm{D|PBH}/\rm{E}}$ determine the regime of anomalous lensing. In the case of $\Xi_{\rm{D|PBH}/\rm{E}}>1$, the Dyson sphere\textendash like obscure the lens effect. Otherwise, we will have an obscuration of one or none of the images. It is intriguing to note that this ratio is independent of the mass of the PBH. It only depends on the configuration of the observer, lens and source. It also depends on the characteristics of the Dyson sphere\textendash like structure.
For the energy efficiency factor of $\epsilon=0.2$, the Eddington ratio of $\eta=10^{-4}$, $D_s \simeq 8 \rm{kpc}$ and $x=0.5$, the critical temperature which gives $\Xi_{\rm{D|PBH}/\rm{E}}=1$ is $T_c \simeq 245$ K. For $T>T_c$, the Dyson sphere partially obscures the hypothetical surface with Einstein radius. 
We model the photometric effects of Dyson sphere\textendash like structures surrounding PBHs by modifying the standard microlensing amplification function. 
This chromatic signature\textemdash suppression in optical microlensing signal and enhancement in infrared (IR)\textemdash provides a distinct observational signature of an energy-harvesting megastructure. To model a more realistic Dyson-sphere\textendash like structure, we assume that the Dyson sphere\textendash like structure is a swarm \citep{smkith2022} with optical depth of $\tau_{\rm{SWARM}} = 0.2$ with area coverage of $10\%$. In the next section, we will present our results.
 
\begin{figure}
	\centering
	\includegraphics[scale=0.30]{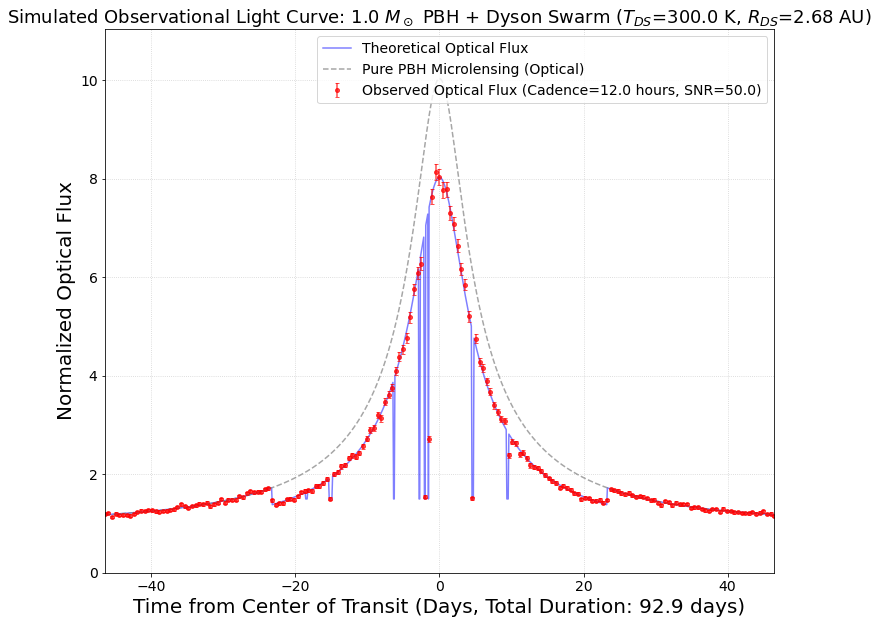}
	\caption{Microlensing light-curve of a solar mass PBH with Dyson sphere\textendash like structure (swarm) is plotted versus time in days. The temperature and radius of this structure are assigned to be $T_{\rm{DS}}=300~ \rm{K}$ and $R_{\rm{DS}}= 2.68~ {\rm{au}}$. We assume an optical depth of $\tau_{\rm{Dyson}}=0.2$ with $10$ flickers.} \label{fig:1}
\end{figure}

\section{Results and Observational Prospects}
\label{Sec4}
In this section, we present our results. In the first subsection, we show the results of the anomalous light curve. In the second subsection, we discuss the characteristics of long-wavelength excess. In the third subsection, we calculate the optical depth of the microlensing effect in a configuration discussed in this work.
\subsection{Anomalous Light Curve of Microlensing}
For lensing configuration, as a proof of concept, we choose $D_l=4~ {\rm{kpc}}$, $D_s=8 ~{\rm{kpc}}$, $v_{\rm{T}}=200 ~ \rm{km/s}$ and $u_0=0.1$. The typical velocities of stars in the disk are in order of $\sim 200 $ \rm{km/s} and the more efficient lensing configuration is when the lens is midway between the midpoint of source and observer. The  $\sim 8 ~{\rm{kpc}}$ is the scale of the distance of the sun to the centre of the galaxy and almost to the edge of the disk. Accordingly, the parameters are chosen as typical of the case.\\
\begin{figure} 
	\centering
	\includegraphics[scale=0.30]{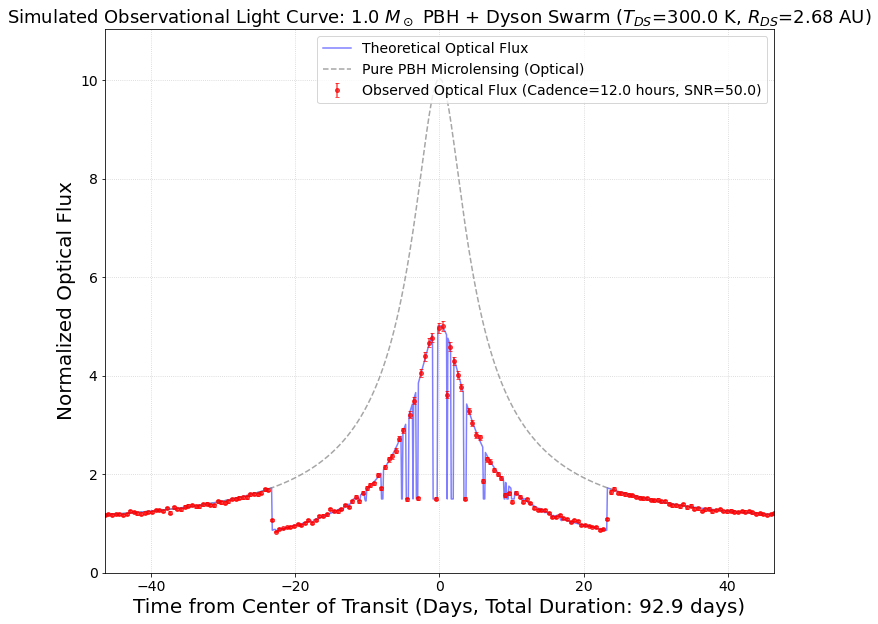}
	\caption{Microlensing light-curve of a solar mass PBH with Dyson sphere\textendash like structure (swarm) is plotted versus time in days. The temperature and radius of this structure are assigned to be $T_{\rm{DS}}=300~ \rm{K}$ and $R_{\rm{DS}}= 2.68~ {\rm{au}}$. We assume an optical depth of $\tau_{\rm{Dyson}}=0.5$ with $50$ flickers.} \label{fig:2}
\end{figure}
In Figure \ref{fig:1}, we plot the normalized optical flux during a microlensing event in terms of time (days).
The gray dashed line shows the theoretical Pacynski microlensing light curve for a solar mass PBH. The blue anomalous light curve is a simulated light curve of PBH with a Dyson swarm in $T_{\rm{Dyson}}=300.0~ \rm{K}$ and radius of $R_{\rm{Dyson}}|_{\rm{\tiny{PBH}}}=2.68~ \rm{au}$ (For brevity we use the subscript "DS" in Figure \ref{fig:1} to indicate Dyson sphere). Due to the optical depth of the Dyson swarm $\tau_{\rm{Dyson}} \sim 0.2$, the amplitude of magnification is dimmed. In this figure we depicted the effect of random $10$ flickers with a duration time of $10^{-3}\times t_{\rm{E}} ~ {\rm{up~ to}}~ 10^{-2}\times t_{\rm{E}}$. The amplitude change is randomly chosen  with $\sim \pm 5\%$ of the magnified amplitude.\\
Ordinary PBH microlensing lacks flickers; their presence hints at structured material (e.g., a swarm).
\begin{figure}
	\centering
	\includegraphics[scale=0.30]{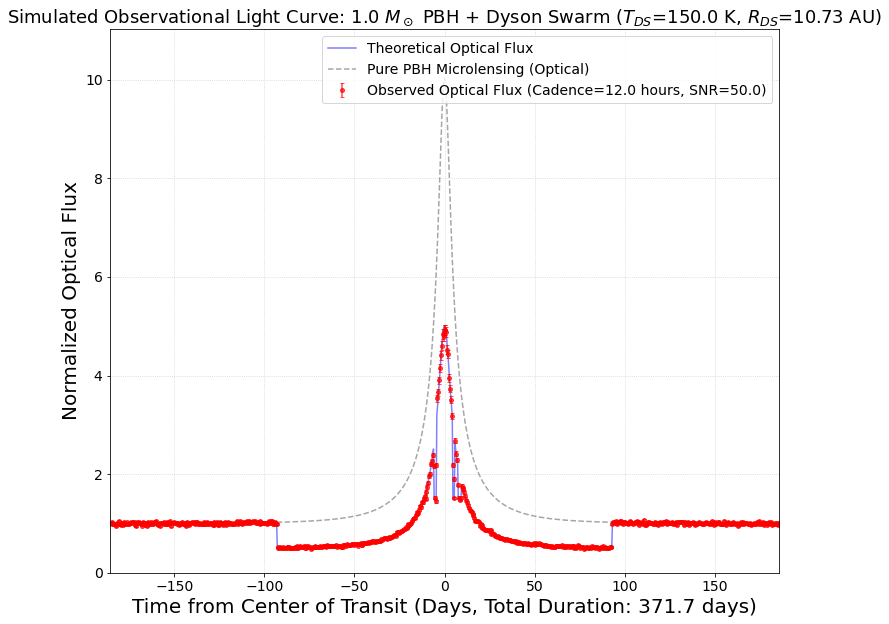}
	\caption{Microlensing light-curve of a solar mass PBH with Dyson sphere\textendash like structure (swarm) is plotted versus time in days. The temperature and radius of this structure are assigned to be $T_{\rm{DS}}=150~ \rm{K}$ and $R_{\rm{DS}}= 10.73~ {\rm{au}}$. We assume an optical depth of $\tau_{\rm{Dyson}}=0.5$ with $50$ flickers.} \label{fig:3}
\end{figure}

 {{A real Dyson swarm would not be a uniform smooth shell — it would be a swarm of many orbiting structures. Depending on the alignment at a given moment, the net effect could be either: dimming events (obscuration/absorption), or brightening events (reflection/glint). These produce a light curve that "flickers" randomly around the smooth theoretical microlensing curve. The shown asymmetry, which means more dimming than brightening is due to the fact that $\tau_{\rm{Dyson}}= 0.2$ already reduces baseline flux during transit.}} These flickers can be accounted for as an observational fingerprint of Dyson swarms. \\
For the case of clarity and comparison in Figure \ref{fig:2}, we plot the same configuration with an optical depth of $\tau_{\rm{Dyson}}=0.5$ and $50$ flickers. {{We should note that we introduced an opaque/semi-transparent swarm at radius $R_{\rm{DS}}$.
{{As the background source star moves across the lens plane, its apparent position may pass within the projected angular radius of the PBH’s surrounding Dyson swarm. Whenever this occurs, the observed (microlensed) flux is reduced by approximately $(1-\tau _{\rm{Dyson}})$ relative to the standard curve; outside the swarm, the suppression vanishes. The light curve therefore shows a step-like drop at ingress (when the source path enters the swarm’s boundary) and a corresponding rise at egress (when it exits). In other words, when the source star’s apparent trajectory (due to the lens–source transverse motion) lies within the angular extent of the PBH’s Dyson swarm, the observed flux is multiplied by roughly $(1 – \tau_{\rm{Dyson}})$.}}\\
In Figure \ref{fig:3}, we plot the configuration where we set the $T_{\rm{DS}}=150~\rm{K}$ and $R_{\rm{DS}}= 10.73~ {\rm{au}}$. We assume an optical depth of $\tau_{\rm{Dyson}}=0.5$ with $50$ flickers. Due to the larger radius of the Dyson sphere in comparison to the Einstein radius, the obscuration is the dominant feature of the effect. In Figure \ref{fig:4}, another extreme is shown. we plot the configuration where we set the $T_{\rm{DS}}=900~\rm{K}$ and $R_{\rm{DS}}= 0.30~ {\rm{au}}$. We assume an optical depth of $\tau_{\rm{Dyson}}=0.5$ with $50$ flickers. }}

\begin{figure}[t!]
	\centering
	\includegraphics[scale=0.30]{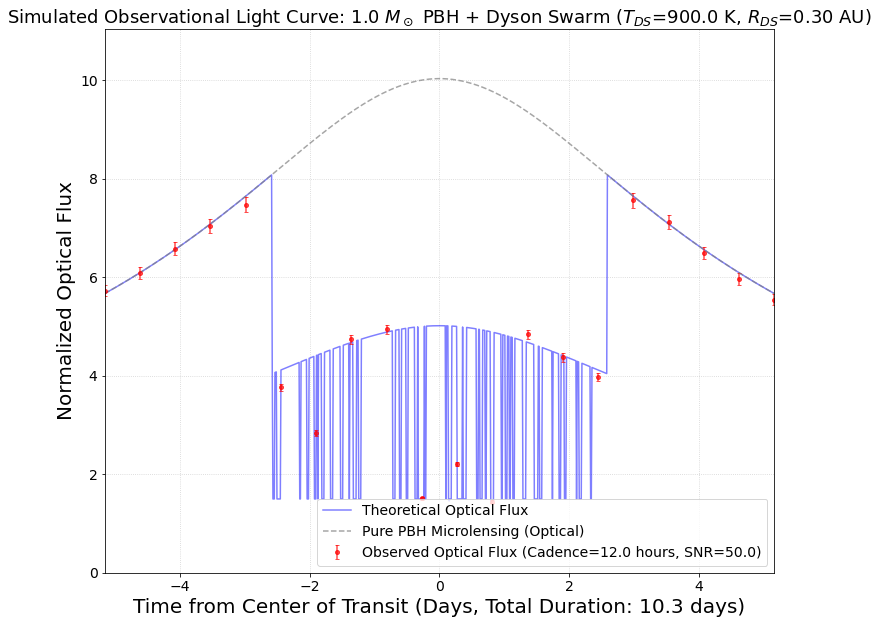}
	\caption{Microlensing light-curve of a solar mass PBH with Dyson sphere\textendash like structure (swarm) is plotted versus time in days. The temperature and radius of this structure are assigned to be $T_{\rm{DS}}=900~\rm{K}$ and $R_{\rm{DS}}= 0.30~ {\rm{au}}$. We assume an optical depth of $\tau_{\rm{Dyson}}=0.5$ with $50$ flickers.} \label{fig:4}
\end{figure}
For observing the flickers, we require high-cadence monitoring (like Vera Rubin Observatory \citep{2022ApJ...928....1W}) as a strategy and observing plan. Flicker's duration and amplitude could reveal
the geometry of the swarm, the size and distribution of swarm components.
{{In Figures \ref{fig:1} to \ref{fig:4}, red dotted points are simulated data of a mock observation with a span of $0.5 ~\rm{day}$ and with detection criteria of signal to noise (SNR) of $50$. SNR is manually set as a fixed value, for proof of concept. Source brightness (e.g., AB magnitude or flux density), instrumental throughput (efficiency, filter bandpass), telescope aperture, exposure time, background noise, read noise, dark current, etc., affect this value. 
We can adopt signature-dependent SNR targets. For broadband IR–optical excess and $10–20\%$ baseline suppression, ${\rm{SNR}} \approx 10–15$ per epoch suffices for model discrimination with multiple in-swarm data points. SNR $\gtrsim 30–50$ is only required if we aim to resolve individual $\pm5\%$ ‘flickers’ at high cadence (durations $\sim 10^{-3}–10^{-2} t_{\rm{E}}$).
{{For a specific example, for JWST/NIRCam at $3.56$\,$\mu\mathrm{m}$ (the F356W filter),  detecting a faint source of magnitude  $m_{\rm{AB}}=28.7$ at ${\rm{SNR}}=10$ requires an exposure time of  $\simeq 10 ~{\rm{ks}}$. This corresponds to a $1\sigma$ flux of $\sigma_1\simeq 1.2 ~n {\rm{Jy}}$.   Note that the flux in Jansky ($1{\rm{Jy}} = 10^{-23} {\rm{erg}}~ s^{-1}{\rm{cm}}^{-2}{\rm{Hz}}^{-1}$) is related to AB magnitude as $F[{\rm{Jy}}]=3631 \times 10^{-0.4m_{\rm{AB}}}$ and, under simple assumptions, the exposure time scales as $t_{\rm{exp}} = t_1(\frac{\rm{SNR}\times \sigma_1}{F})^2$, where $t_1$ and $\sigma_1$ are a reference exposure time and $1\sigma$ flux uncertainty. Accordingly, we can estimate the SNR of observation with JWST/NIRCam as a function of exposure time and apparent magnitude as ${\rm{SNR}}(m,t_{\rm{exp}})=10\times 10^{-0.4 (m-28.7)}\sqrt{{t_{\rm{exp}}}/{( 10~ ks)}}$. This implies that for a source of $m=23.5$ with $t_{\rm{exp}}\simeq 100~s$ we achieve ${\rm{SNR}}\sim 120$   \citep{2016SPIE.9910E..16P}. These values are consistent with the JWST Exposure Time Calculator (ETC) for NIRCam/F356W in a typical low-background field. 
The complete modelling of SNR can be a specifically valuable side project}}\footnote{One can use the online JWST Exposure Time Calculator: \\ \url{https://jwst.etc.stsci.edu/}}.}} \\
In order to check the plausibility of the detection, we have to compare two models. Due to the null hypothesis of the standard microlensing model with no modulation, 
the alternative hypothesis of microlensing plus the  Dyson swarm modulation (e.g., periodic dips, stochastic attenuation, wavelength-dependent suppression). The chi-squared analysis will show that the modified light curve has a reduced chi-square. \\ 
In Figure \ref{fig:5}, the heatmap of optical transit duration of Dyson swarm in days is plotted versus the mass range of PBH in the scale of $0.1-100 M_{\odot}$ and Dyson swarm temperature. The Optical transit duration is calculated from $	(R_{\rm{Dyson}}|_{\rm{\tiny{PBH}}})/v_{\rm{T}}$, where $	R_{\rm{Dyson}}|_{\rm{\tiny{PBH}}}$ is obtained from Equation (\ref{eq:Rdyson-pbh}). Optical transit time is comparably large for massive PBH with low temperature Dyson swarms. In the next subsection, we discuss the long-wavelength excess as a complementary observation.
\begin{figure}
	\centering
	\includegraphics[scale=0.40]{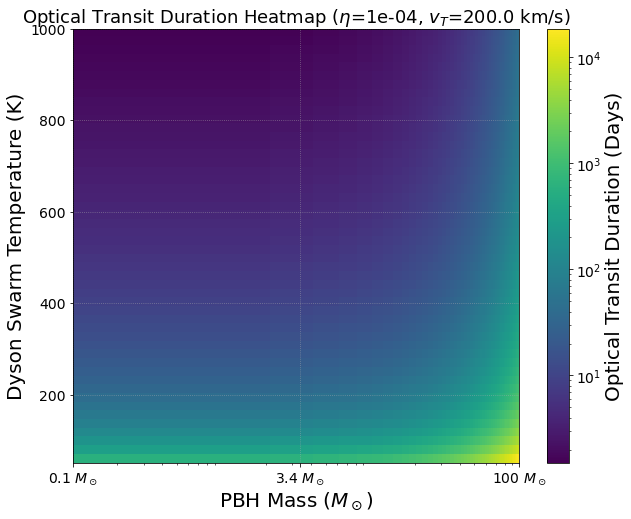}
	\caption{The heatmap of optical transit duration of Dyson sphere\textendash like structure in days plotted versus the mass range of PBH in scale of $0.1-100 ~M_{\odot}$ and Dyson swarm temperature with Eddington ratio $\eta=10^{-4}$, efficiency $\epsilon=0.2$ and transverse velocity of $v_{\rm{T}}=200 ~\rm{km/s}$. } \label{fig:5}
\end{figure}

\subsection{long wavelength excess}
Except for the microlensing amplification, one can anticipate an excess in long wavelengths, which is a motivation for the search for this signature. \citep{1959Natur.184..844C,2022MNRAS.512.2988S}. This excess is depicted in Table \ref{Table1}. {{The luminosity of the Dyson swarm megastructure is
\begin{equation}
	L_{\rm{DS}}=(1-\epsilon)f_{\rm{cap}}L_{\rm{source}}=4\pi \Upsilon R^2_{\rm{Dyson}|{\rm{\tiny{PBH}}}}\sigma_{\rm{SB}} T_{\rm{Dyson}}^4,
\end{equation}
where $\Upsilon$ account for the radiating fraction. The flux received at Earth is $F_{\rm{DS}|\Earth}= 	L_{\rm{DS}} / (4\pi D_{l}^2)$.}} \\
In this calculation, we assume a perfect blackbody emission, no absorption by the interstellar medium and a face-on observation geometry. The bolometric flux of the Dyson Swarm at Earth depends linearly on PBH mass ($F_{\rm{DS}|\Earth}\propto M_{\rm{PBH}}$), and is independent of the swarm temperature, since the radius–temperature scaling cancels.
The normalization is arranged as follows: We set the detection limit of James Webb Space Telescope (JWST \footnote{JWST is designed to observe light from the infrared portion of the electromagnetic spectrum, ranging from 0.6 to 28.5 microns.}), which is $0.1~ \mu \rm{Jy}$ in $10~\mu \rm{m}$ \citep{2023PASP..135d8003W}. {{Also, note that JWST's best sensitivity is in NIRCam (0.6-5 $\mu$m). Imaging alone is a first step to detect/differentiate Dyson swarm signatures, and spectroscopy will be a complementary tool for confirmation.}} 
\begin{figure}[t]
	\centering
	\includegraphics[scale=0.35]{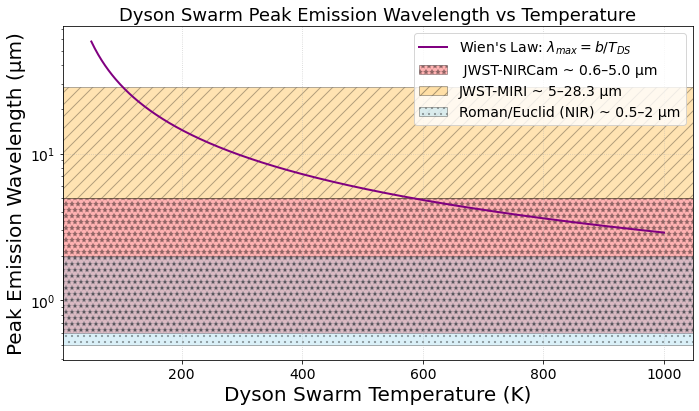}
	\caption{Peak emission wavelength of a Dyson Swarm as a function of swarm
		temperature, following Wien’s displacement law,
		\(\lambda_{\max} = b/T_{\rm DS}\). The shaded regions indicate the
		approximate wavelength coverage of key infrared observatories: JWST/NIRCam (red band with star hatching, \(0.6\!-\!5.0\,\mu\mathrm{m}\) )
		JWST/MIRI (orange band with slanted hatching, \(5.0\!-\!28.3\,\mu\mathrm{m}\)) and
		Roman/Euclid (NIR) (blue band with dotted hatching, \(0.5\!-\!2.0\,\mu\mathrm{m}\)).
		For example, a swarm at \(T_{\rm DS}\!\sim\!300\,\mathrm{K}\) peaks near
		\(\sim 10\,\mu\mathrm{m}\), within JWST- MIRI’s sensitivity, whereas hotter swarms
		(\(T_{\rm DS}\gtrsim1000\,\mathrm{K}\)) peak in the near-IR range of Euclid and
		Roman.} \label{fig:6}
\end{figure}
\begin{figure}[t]
	\centering
	\includegraphics[scale=0.37]{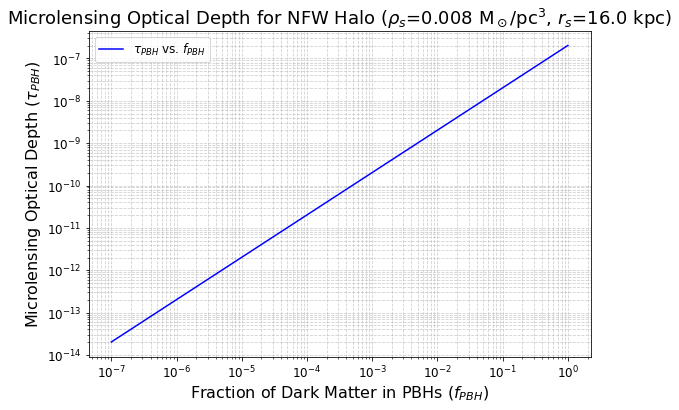}
	\caption{Microlensing optical depth is plotted versus fraction of DM as PBH for a halo profile of NFW with $\rho_s = 0.008 M_{\odot} / {\rm{pc}}^3$ and $r_s=16.0 \rm{kpc}$. } \label{fig:7}
\end{figure}
The peak emission wavelength of a Dyson Swarm follows Wien’s displacement law,
\(\lambda_{\max} = b/T_{\rm DS}\), and is therefore determined solely by the swarm
temperature. Figure \ref{fig:6} illustrates this relation, with shaded bands indicating
the wavelength coverage of major IR facilities. For example, swarms with
\(T_{\rm DS} \sim 300\,\mathrm{K}\) peak at \(\lambda_{\max} \approx 10\,\mu\mathrm{m}\),
squarely within JWST/MIRI’s sensitivity range, while hotter swarms
(\(T_{\rm DS} \gtrsim 1000\,\mathrm{K}\)) shift into the near-IR regime accessible to
Euclid and Roman. This representation emphasizes the observational relevance of
different Dyson Swarm temperatures and avoids redundant dependence on PBH mass,
which cancels in this relation.
\subsection{Optical Depth of the Microlensing Effect}
To estimate the detection probability and the likelihood of finding technosignatures via microlensing, an important quantity to compute is the optical depth of the lensing. This parameter is defined due to Equation (\ref{eq:tau-rho}) as
\begin{equation}
	\tau _{\rm{PBH}} = \int_0^{D_s}\frac{4\pi G}{c^2} \rho_{\rm{PBH}}(D_l)\frac{D_l(D_s-D_l)}{D_s}dD_l, \label{eq:tau}
\end{equation}
where $D_l$ is the distance to the lens, $D_s$ is distance to the source and $\rho_{\rm{PBH}}$ is PBH density at $D_l$. Note that 
\begin{equation}\label{eq:rhoDl}
	\rho(D_L) = f_{\rm{PBH}} \rho_{\rm{DM}}(D_L), 
\end{equation} 
where $ f_{\rm{PBH}}$ is the fraction DM made of PBH.
For DM halo profile we use the Navarro-Frenk-White (NFW) \citep{1997ApJ...490..493N}
\begin{equation}
	\rho_{\rm{DM}}(r)=\frac{\rho_s}{({r}/{r_s}) \left(1+{r}/{r_s}\right)^2},
\end{equation}
where $\rho_s$ and $r_s$ are the specific density and radius of the NFW profile. For Milky way galaxy, We choose $\rho_s=0.008 M_{\odot}/{\rm{pc}}^3$ and $r_s = 16 ~\rm{kpc}$.
We set the position of observer in $\sim 8~ \rm{kpc}$ of the centre of the galaxy and we insert the NFW profile in Equations (\ref{eq:tau} and \ref{eq:rhoDl}). The free parameter of the optical depth is $f_{\rm{PBH}}$. The optical depth of detecting a PBH lens versus $f_{\rm{PBH}}$ is plotted in Figure \ref{fig:7}. 
For a lensing configuration with Dyson sphere\textendash like structure, $\tau_{\rm{Dyson}}$,  we will have
\begin{equation}
	\tau_{\rm{Dyson}}=f_{\rm{Dyson}} \times 	\tau _{\rm{PBH}},
\end{equation}
where $f_{\rm{Dyson}}$ is the fraction of PBH that are harvested by advanced civilization. The observational lensing rate of anomalous (non-Paczynski) lightcurve is
\begin{equation}
	\Gamma_{\rm{Dyson}} \simeq  \frac{	\tau_{\rm{Dyson}}}{\langle t_E \rangle},
\end{equation}
where  $\langle t_{\rm{E}} \rangle \simeq  \frac{ \langle R_{\rm{E}} \rangle}{\langle v_{\perp}\rangle}$ is the averaged Einstein crossing time, which depends on masses and velocity distribution of lenses.
If we assume a conservative fraction of $f_{\rm{PBH}}=0.01$ and $f_{\rm{Dyson}}=0.01$, the typical microlensing optical depth to the center of galaxy is roughly $\sim 10^{-6}$, implying $\tau_{\rm{PBH}} \sim 10^{-8}$, which is an almost agreement with Figure \ref{fig:7}. Accordingly,  $\tau_{\rm{Dyson}}\sim 10^{-10}$.\\
The observational lensing rate for a typical relative velocity of $\langle v_{\perp} \rangle \sim 200~ \rm{km/s}$ and solar mass PBH in configuration of $x=0.5$ will be $\Gamma_{\rm{Dyson}}\simeq 10^{-9} \rm{star}^{-1} \rm{year}^{-1}$.
This is an achievable rate for upcoming microlensing surveys.
\section{Conclusions and Future Remarks} 
\label{Sec5}
The prospect of detecting extraterrestrial technosignatures via microlensing opens a new interdisciplinary frontier, bridging astrophysics, cosmology, SETI and search for
extraterrestrial artificial intelligence (SET-AI) \citep{2025ApJ...978..132B} type research. Our study has demonstrated that even stochastic, partially absorbing Dyson sphere\textendash like structures, like Dyson Swarms around PBHs, can leave detectable imprints in time-domain surveys such as James Webb Space Telescope, the Nancy Grace Roman Space Telescope \citep{2022ApJ...928....1W} and the Vera Rubin Observatory \citep{2023arXiv230613792S}. \\
In this work, we demonstrate that the anomalous microlensing signal and the excess infrared radiation are complementary probes for searching for technosignatures. We also compute the optical depth and event rate of this type of proposal. {{Additionally, under conservative assumptions, we conclude that for an effective one-year observation of approximately $\sim 10^{9}$ stars in our galaxy, there exists a chance of detecting a Dyson sphere\textendash like structure in this context. However, our understanding of the probability that Dyson spheres exist remains very uncertain, and this conclusion should be viewed with caution.}}
For our proposal, several avenues for future investigation are compelling, such as Bayesian inference and model selection for the detection of the microlensing effect. Also, for future studies, one can take into account the wavelength-dependent detector responses and the possible obscuration by the swarm itself and also the time variability of the emission. \\
Chromatic and polarization signatures are also other paths of investigation which can be induced by engineered materials to enrich the detectability of artificial swarms.\\
One can also use the machine learning classifiers to train supervised learning models on synthetic datasets of modulated microlensing light curves to enhance detection efficiency.\\
Microlensing parallax and high cadence follow-up are other paths of research that one can follow. Parallax effects can be used to break degeneracies and improve parameter estimation. 
As time-domain astronomy enters an era of unprecedented sensitivity and sky coverage, these proposed directions will refine the search for technosignatures and find the signature of advanced civilizations.
	\section*{ACKNOWLEDGMENTS}
	We thank the anonymous referee, whose insightful comments and detailed suggestions elevated the manuscript to a new level.\\
SB is partially supported by the Abdus Salam International Center for Theoretical Physics (ICTP) under the regular associateship scheme.
SB is partially supported by the Sharif University of Technology Office of the Vice President for Research under Grant No. G4010204.

	\iftrue
	
	\fi
	
\end{document}